\documentclass[10pt,twocolumn, showpacs]{revtex4-1}

\usepackage{times}
\usepackage{epsfig,wrapfig}
\usepackage{epstopdf} 
\usepackage{amssymb}
\usepackage{amsmath}
\usepackage{fancyhdr}
\usepackage{subfigure}
\usepackage{color}

\begin{document}

\title{Extreme optical activity and circular dichroism of chiral metal hole arrays}

\author{M.V. Gorkunov$^1$, A.A. Ezhov$^{2,3}$, V.V. Artemov$^1$, O.Y. Rogov$^1$, and S.G. Yudin$^1$}
\affiliation{$^1$A.V. Shubnikov Institute of Crystallography, Russian Academy of Sciences, 119333 Moscow, Russia\\
$^2$M.V. Lomonosov Moscow State University, 119991 Moscow, Russia\\
$^3$A.V. Topchiev Institute of Petrochemical Synthesis, 119991 Moscow, Russia}

\begin{abstract}

We report extremely strong optical activity and circular dichroism exhibited by subwavelength arrays of four-start-screw holes fabricated with one-pass focused ion beam milling of freely suspended silver films. Having the fourth order rotational symmetry, the structures exhibit the polarization rotation up to 90 degrees and peaks of full circular dichroism and operate as circular polarizers within certain ranges of wavelengths in the visible. We discuss the observations on the basis of general principles (symmetry, reciprocity and reversibility) and conclude that the extreme optical chirality is determined by the chiral localized plasmonic resonances.
\end{abstract}

\maketitle

Structuring of metals and semiconductors at the scales of tens and hundreds of nanometers gives rise to fundamentally new functional optical properties inaccessible with conventional materials \cite{Schuller}. Apart from the anticipated revolutionary applications for superlenses and optical camouflage \cite{Shalaevbook}, such materials have been confirmed to perform the optical spectral and polarization filtering \cite{filtering}, amplify the stimulated emission \cite{Stockman}, provide efficient gas- \cite{Giessen} and bio-sensing \cite{Brolo}, etc.

One can efficiently tailor the optical properties of the structures by varying shape and size of the structural elements, while the formal chemical composition is less important. A lot of attention has been given to the structured materials with enhanced optical chirality \cite{twist}. Since the observable effects of optical chirality are very weak in conventional media, fabrication of composites capable of substantial polarization rotation and circular dichroism (CD) is desirable for the advanced optical technologies.

Despite the wide range of emerging fabrication techniques, obtaining chiral structures with optically subwavelength periods remains challenging. Relatively simple and widely used planar nanofabrication approaches such as the electron lithography are limited to the structures with the so-called 2D-chirality \cite{quasi2D, Plum}. However, the absence of true chirality leads to a noticeable lack of optical activity and CD \cite{quasi2D}, which may be mimicked by planar structures only in conjunction with strong birefringence \cite{Plum}. Using multistage layer-by-layer processing one can produce micron-scale truly chiral structures suitable for the infrared wavelengths of a few microns \cite{Decker}. Similarly, by the direct laser writing a bulk polymer photonic crystal with pronounced chiral transmission selectivity in the infrared has been fabricated \cite{beamsplitter}, however, the low spatial resolution prevents from fabricating the structures subwavelength in the visible. Another fairly complex method of glancing angle deposition has been used for the formation of nanoscale dielectric helical templates for the plasmonic nanoparticle decoration \cite{Singh} and metallic helices \cite{Gibbs}, which have shown a noticeable CD of the order of several degrees and significant extinction.

In this Letter we report an alternative concept of optical chiral materials based on light transmitting chiral hole arrays. Hole and slit arrays exhibit the extraordinary optical transmission (EOT) in the subwavelength regime \cite{Ebbessen}. It has been shown that arrays of one-start screw holes in metal screen provide CD in the THz range \cite{THz}. Numerical simulation of light transmission through chiral holes has shown the possibility of CD in the visible \cite{Ezhov}.
We demonstrate that imparting chiral shape to nano-size holes results in extreme values of optical chirality: the peaks of full CD in the visible when the structures operate as circular polarizers and the polarization rotation angle reaches 90$\rm ^o$.
Notably, one can conveniently fabricate such structures using the common single-pass focused ion beam (FIB) technique with varying etching time.
\begin{figure}
\centering
\includegraphics[width=7cm]{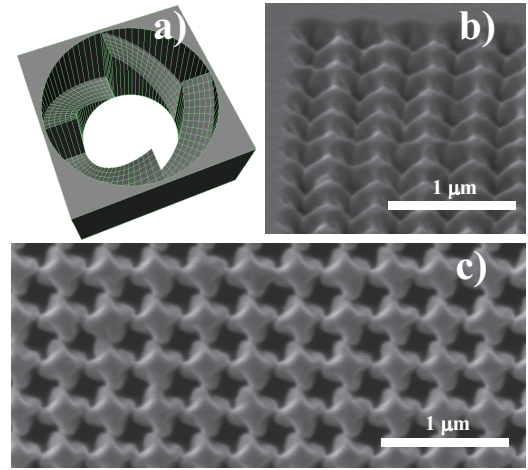} \caption{(a): 3D-model of the structure unit cell as implemented into the FIB milling digital template. SEM images of parts of the structure milled in a single pass in 380 nm thick silver foil: tilted by 52$\rm ^o$ (b) and normal (c) views.}\label{fig1}
\end{figure}

The chiral holes have been milled in freely suspended silver foils. The foils have been prepared by vacuum silver sputtering on potassium hydrogen phthalate substrates, which have been afterwards dissolved in water. The foils were stable enough to be suspended upon a circular window of 1.5 mm in diameter. The FEI Helios DualBeam microscope has been used for the single-pass processing of the foils with the focused beam of gallium ions of the current of 30 pA and the energy of 30 eV. The ion beam path and dwell time have been set by the digital template (stream file) prepared accordingly to the 3D model of the structure unit cell shown in Fig.~\ref{fig1}a). The unit cell comprises a 4-start screw thread circular hole with the inner diameter of 187 nm and the outer diameter of 337 nm. Square 72$\times$72 arrays of the holes with the period of 375 nm have been milled resulting in 27$\times$27 $\mu$m$^2$ areas sufficient for the optical transmission measurements. Separate empty windows of the same area have also been milled for the optical reference purposes.

As seen in Figs.~\ref{fig1} b) and c), the resulting shape of the structure differs from the milling template due to the inevitable ion beam defocusing and diversion by the curved silver surface. At the same time, the screw symmetry breaking important for the structural chirality, i.e., the absence of mirror planes, is clearly present. Two samples of the same type and equal lateral dimensions have been milled in the foils of different thickness: 270 nm (sample S1) and 380 nm (sample S2).

\begin{figure}
\centering
\includegraphics[width=8cm]{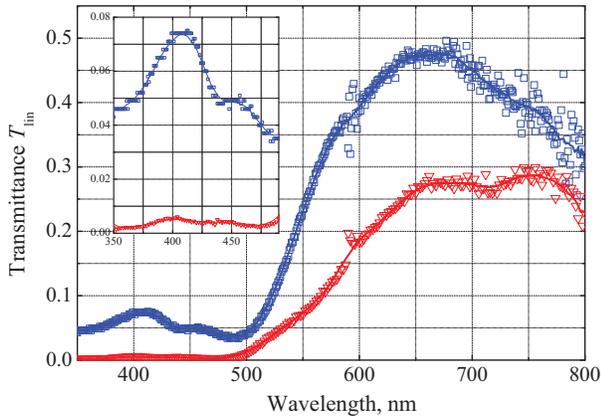}
\caption{Measured transmittance spectra for linearly polarized light incident on S1 (squares) and S2 (triangles)  structures. The inset shows the short wavelength range in more detail. Solid lines are drawn to guide the eye.}\label{fig2}
\end{figure}

Microspectroscopy and microspectropolarimetry of light transmitted through the samples was carried out with a spectroscopic Uvisel 2 (Horiba Jobin-Yvon) ellipsometer  capable of measurements for the normal incidence with the light spot size of 100 $\mu m$ sufficiently small for studying the individual arrays. The incident light was linearly polarized. Measured transmission spectra were normalized by the reference window transmission.
As seen in Fig.~\ref{fig2}, the structures possess a broad transmission band in the red part of the spectrum.

For the linearly polarized input beam, the optical chirality can be sufficiently characterized by the two parameters of the output elliptically polarized light: the rotation angle $\Phi$ defined as the angle between the direction of the incident polarization and the direction of the long main axis of the polarization ellipse of transmitted light, and the ellipticity $\Psi$ defined as the arctangent of the ratio of main polarization ellipse axes. The latter can be conveniently reduced to CD as $D=\sin2\Psi$, which characterizes the difference in transmittance of left and right polarized light: $D=(T_R-T_L)/(T_R+T_L)$.

To ensure that the samples possess exactly circular optical anisotropy, the results for different directions of the incident light polarization have been compared. Also the measurements have been performed for light incident on different sides of the samples. For both samples the ellipsometry data have shown very weak dependence on the incident polarization as well as on the side of incidence.

\begin{figure}
\subfigure{\includegraphics[width=8cm]{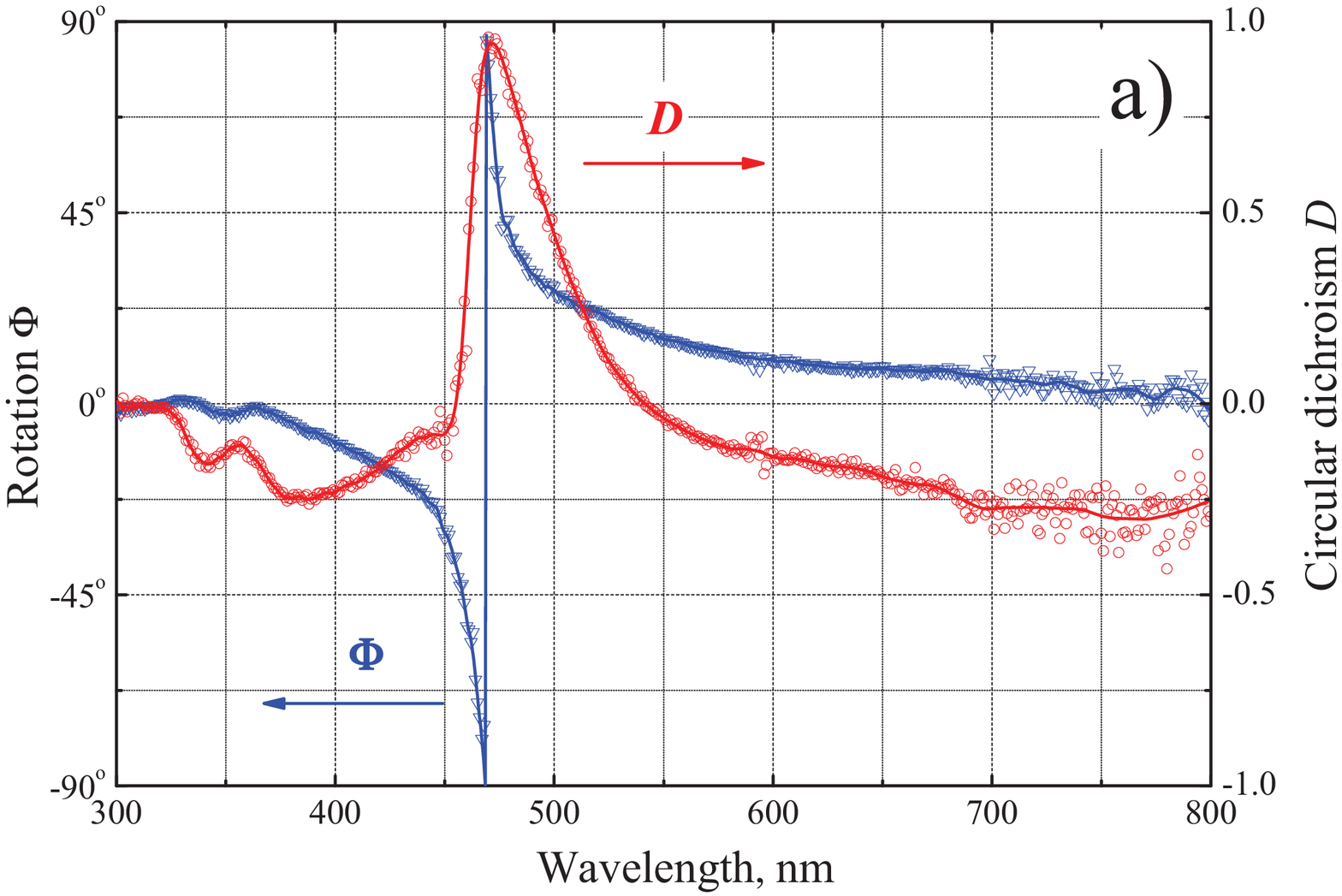}}\\
\subfigure{\includegraphics[width=8cm]{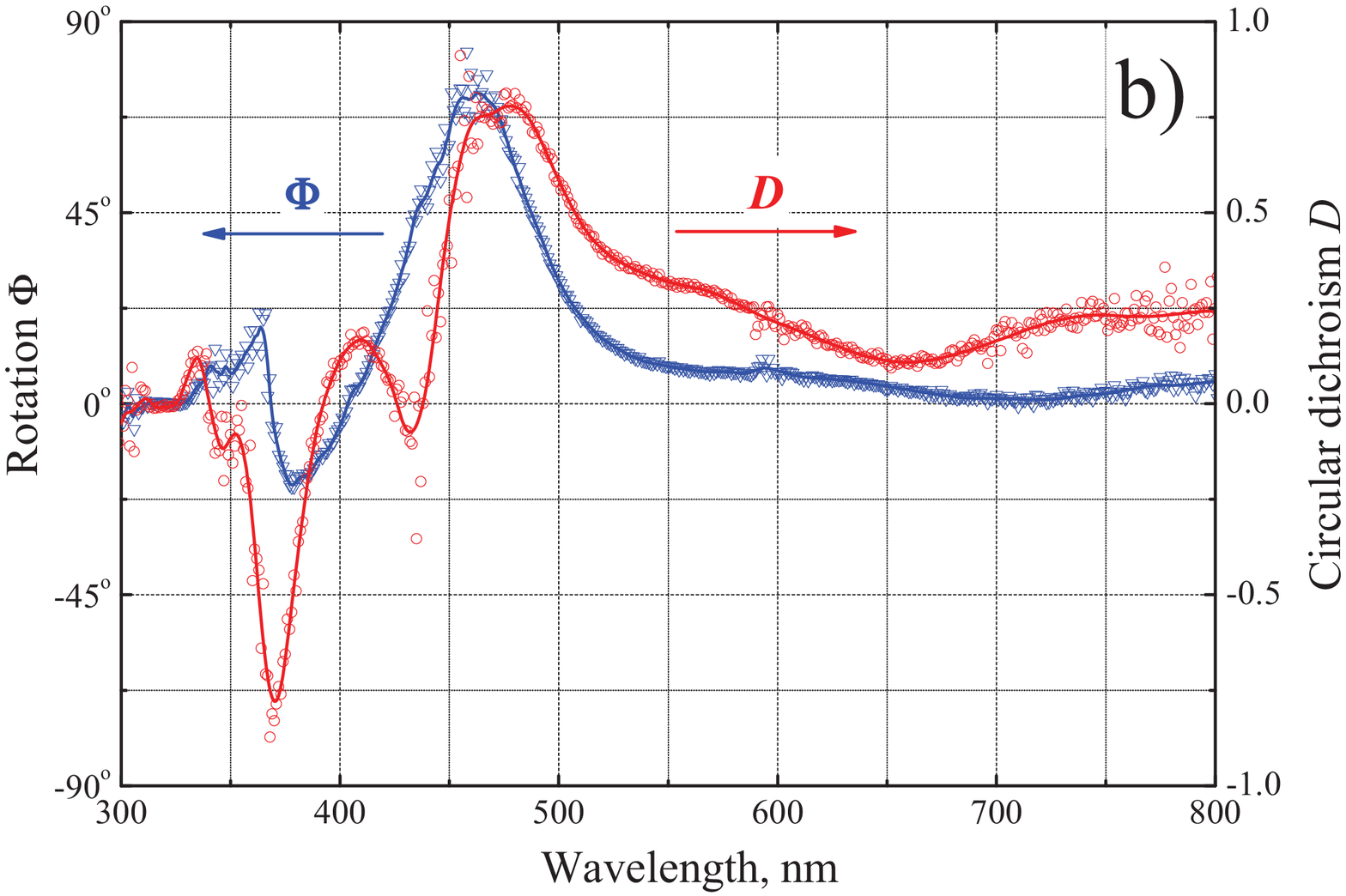}}
\caption{Measured optical activity (triangles) and circular dichroism (squares) for the samples S1 (a) and  S2 (b). Solid lines are drawn to guide the eye.}\label{fig3}
\end{figure}

Typical measured spectra of $\Phi$ and $D$ for both samples are shown in Fig.~\ref{fig3}. Note that the scale of vertical axes of the plots covers the whole range of possible values of $\Phi$ and $D$. Thus the fabricated structures truly exhibit extreme optical chirality in the visible. In both samples, the maximum chirality occurs around the 470 nm wavelength, where the output beam is circularly polarized, i.e., the structures work as full circular polarizers. At larger wavelengths, the output polarization approaches linear and the rotation angle decreases gradually from the extreme values to almost zero in the near infrared.

At shorter wavelengths, the difference in sample thickness affects the spectra critically. While the thinner sample S1 shows only very moderate optical chirality at wavelengths close to the structure period of 375 nm (the diffraction Wood-Rayleigh anomaly), for the thicker sample S2 the anomaly causes an almost extreme peak of CD of inverse handedness accompanied by a considerable optical activity.

Obviously, the known simple models of light transmission through holes and slits (see e.g. Refs. \cite{theorholes,Sturman}) are not applicable to such complex chiral structures. However, relying on fundamental principles of spatial symmetry, reciprocity and reversibility one can advance substantially in understanding the physics beyond the observed optical properties.

\begin{figure}
\includegraphics[width=6cm]{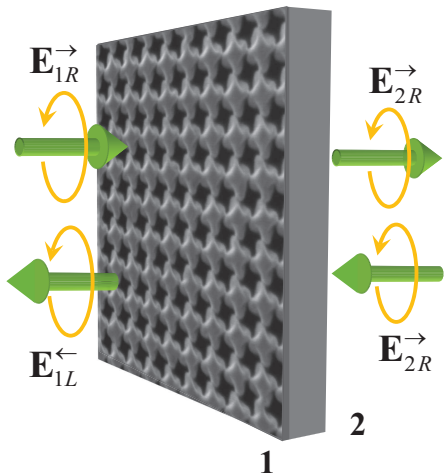}
\caption{Schematic of the general S-matrix transmission-reflection problem.}\label{fig4}
\end{figure}

First, since the linear relation between the input and output two-component complex polarization vectors can be expressed by a 2nd rank tensor, in accordance with Hermann's theorem \cite{Hermann} the 4-fold rotational point symmetry ensures the effective in-plane isotropy of the electromagnetic response. However, the absence of an in-plane mirror symmetry, prevents the response from the total isotropization: The right and left polarized states are the eigenmodes but not necessarily their linear combination is. Accordingly, the enclosed transmission/reflection problem involves circularly polarized beams of the opposite handedness propagating in the opposite directions. One of the two symmetric possibilities is shown in Fig.~\ref{fig4}, for which the incident and outgoing amplitudes can be related by the S-matrix equation:
\begin{equation}\label{general}
\left(
\begin{array}{ll}
E_{1L}^\leftarrow \\
E_{2R}^\rightarrow
\end{array}
\right) =
\left(
\begin{array}{cc}
r_{1RL}\ \ t_{L}^\leftarrow\\
t_R^\rightarrow \ \ r_{2LR}
\end{array}
\right)
\left(
\begin{array}{ll}
E_{1R}^\rightarrow\\
E_{2L}^\leftarrow
\end{array}
\right),
\end{equation}
where  $t_{R}^\rightarrow$ and $t_{L}^\leftarrow$ are the right-polarized (RP) forward and left-polarized (LP) backward transmission amplitudes, while $r_{1RL}$ and $r_{2LR}$ are the 1st side RP$\rightarrow$LP and the 2nd side LP$\rightarrow$RP reflection amplitudes, correspondingly.

Next, according to the Lorentz lemma, in the absence of polarization conversion the transmission reciprocity yields
\begin{equation}\label{recipr}
t_{L}^\leftarrow=t_{L}^\rightarrow \ \ \ \text{and} \ \ \ t_{R}^\leftarrow=t_{R}^\rightarrow,
\end{equation}
which explain the independence of the spectropolarimetry results of the side of incidence. Note also that for a planar 2D-chiral structure, i.e., invariant with respect to the reflection about its plane, these equalities lead to the total cancelation of both CD and rotation. Therefore, one needs indeed the true chirality to achieve the substantial optical activity and CD.

With no external magnetic field applied, the time reversal invariance is equivalent to the absence of dissipation. Then the energies flowing to and from the structure should be equal and for the transmission/reflection process depicted in Fig.~\ref{fig4} and described by Eq.~\eqref{general} this means $|E_{1L}^\leftarrow|^2+|E_{2R}^\rightarrow|^2=|E_{1R}^\rightarrow|^2+
|E_{2L}^\leftarrow|^2$ for any combination of the incident waves. This can only be possible if the S-matrix in Eq.~\eqref{general} is unitary and thus  $|t_{L}^\leftarrow|=|t_R^\rightarrow|$. Combined with \eqref{recipr} this yields zero CD and so the latter is determined by the dissipation.

This close connection of CD and dissipation in chiral structures of 4-fold in-plane symmetry is a very general fact that has been recently pointed out also for the arrays of 4-helices \cite{Kaschke}. As we have seen, the observed CD peaks are indeed located in the range of low transmittance, which suggests that the transmission handedness occurs due to the extinction handedness, i.e., due to specific chiral plasmonic resonances. Leaving the particular details of the resonances beyond the scope of this Letter, one can outline the related application prospects. On the one hand, a lossless circular polarizer operating similarly to, say, a linear wire grid polarizer is impossible: A substantial CD is to be always accompanied by losses. On the other hand, the enhancement of local fields chirality \cite{Super} due to the chiral localized plasmonic resonances should be very advantageous for the sensing and photochemical separation of chiral substances and nanoobjects including those of biological nature. The recent studies on the 2D-chiral structures \cite{Hendry} have shown the amplification of the chiral molecular response and it is clear that the use of truly chiral resonant structures should enhance this effect greatly.

In conclusion, we have demonstrated experimentally that the light transmitting subwavelength arrays of 4-start screw holes in  freely suspended silver films exhibit extremely high up to 330 $\rm ^o/\mu m$ rotatory power and full CD in the visible. Applying the general principles of symmetry, reciprocity and reversibility we have related the observed peaks of CD to localized chiral plasmonic resonances.

We are grateful to  A.L. Vasiliev and Shared Research Centers of IC RAS and MSU supported by the Ministry of Education and Science of the Russian Federation for the equipment provided, to V.L. Manomenova for the potassium hydrogen phthalate crystals, and to N.M. Gorkunov for the assistance during sample preparation.
MVG, VVA, OYR, and SGY acknowledge the support from the Russian Foundation for Basic Research (project 13-02-12151 ofi$_ -$m) and the Russian Academy of Sciences (Program 24), AAE acknowledges the support from the Russian Foundation for Basic Research (project 14-03-00737).



\begin{thebibliography}{10}

\bibitem{Schuller}
J.A. Schuller, E.S. Barnard, W. Cai, Y.C. Jun, J.S. White, and M.L. Brongersma, Nature Mater. {\bf 9}, 193 (2010).

\bibitem{Shalaevbook}
W. Cai and V. Shalaev, {\em Optical Metamaterials: Fundamentals and Applications} (Springer, 2009)

\bibitem{filtering}
T. Xu, Y.-K. Wu, X. Luo, and L.J. Guo, Nature Commun. {\bf 1}, 59 (2010).

\bibitem{Stockman}
D.J. Bergman and M.I.  Stockman, Phys. Rev. Lett. {\bf 90}, 027402 (2003).

\bibitem{Giessen}
N. Liu, M.L. Tang, M. Hentschel, H. Giessen and A.P. Alivisatos, Nature Mater. {\bf 10}, 631 (2011).

\bibitem{Brolo}
A.G. Brolo, Nature Phot. {\bf 6}, 709 (2012).

\bibitem{twist}
M. Wegener and S. Linden, Physics {\bf 2}, 3 (2009).

\bibitem{quasi2D}
M. Kuwata-Gonokami, N. Saito, Y. Ino, M. Kauranen, K. Jefimovs, T. Vallius, J. Turunen, and Yu. Svirko, Phys. Rev. Lett. {\bf 95}, 227401 (2005).

\bibitem{Plum}
E. Plum, X.-X. Liu, V.A. Fedotov, Y. Chen, D.P. Tsai, and N.I. Zheludev, Phys. Rev. Lett. {\bf 102}, 113902 (2009).

\bibitem{Decker}
M. Decker, R. Zhao, C. M. Soukoulis, S. Linden, and M. Wegener, Opt. Lett. {\bf 35}, 1593 (2010).

\bibitem{beamsplitter}
M.D. Turner, M. Saba, Q. Zhang, B.P. Cumming, G. E. Schroeder-Turk, and M. Gu, Nature Phot. {\bf 7}, 801, (2013).

\bibitem{Singh}
J.H. Singh, G.Nair, A. Ghosh, and A. Ghosh, Nanoscale {\bf 5}, 7224, (2013).

\bibitem{Gibbs}
J. G. Gibbs, A. G. Mark, S. Eslami, and P. Fischer, Appl. Phys. Lett. {\bf 103}, 213101 (2013).


\bibitem{Ebbessen}
T.W. Ebbesen, H.J. Lezec, H.F. Ghaemi, T. Thio, and
P.A. Wolff, Nature {\bf 391}, 667 (1998).

\bibitem{THz}
F. Miyamaru and M. Hangyo, Appl. Phys. Lett. {\bf 89}, 211105 (2006).

\bibitem{Ezhov}
A.A. Ezhov, Y.A. Ilyushin, and A.A. Fedyanin, Abstracts of XXXI Int. Sci. Conf. Electronics and Nanotechnology, Kyiv, Ukraine, April 12-14, 23 (2011).

\bibitem{theorholes}
H. Liu and Ph. Lalanne, Nature Lett. {\bf 452}, 728 (2008).

\bibitem{Sturman}
B. Sturman, E. Podivilov, and M. Gorkunov, Phys. Rev. B {\bf 77}, 075106 (2008).

\bibitem{Hermann}
C. Hermann, Zs. Kristallog. {\bf 89}, 32 (1934).

\bibitem{Kaschke}
J. Kaschke, J.K. Gansel, and M. Wegener, Opt. Expr. 20, 26012 (2012).

\bibitem{Super}
Y.Tang and A.E. Cohen, Phys. Rev. Lett. {\bf 104}, 163901 (2010).

\bibitem{Hendry}
E. Hendry, T. Carpy, J. Johnston, M. Popland, R.V. Mikhaylovskiy, A. J. Lapthorn, S.M. Kelly, L.D. Barron, N. Gadegaard and M. Kadodwala, Nature Nanotech. {\bf 5}, 785 (2010).

\end{thebibliography}
\end{document}